\documentclass[10pt]{article}
\usepackage{spconf,graphicx}

\title{spoofing attack detection in dynamic channels with imperfect CSI  \\}
\name{Chu Li, Aydin Sezgin}
\address{Ruhr-Universit\"at Bochum,\\
	Email:  \{chu.li, aydin.sezgin\}@rub.de} 
\usepackage{cite}

\usepackage{graphicx}

\graphicspath{{./figures}}
\DeclareGraphicsExtensions{.pdf}
\usepackage[cmex10]{amsmath}
\usepackage{amssymb}
\usepackage{amsthm}
\usepackage{stfloats}

\usepackage{hyperref}

\usepackage{multirow}
\usepackage{bm}
\usepackage{algorithmic}
\usepackage{xcolor}
\usepackage{tikz}
\usetikzlibrary{shapes,arrows,positioning,plotmarks,shadows,calc,matrix,fit,patterns}
\usepackage{pgfplots}
\usepackage{todonotes}
\usepackage{datetime}
\usepackage{cancel}
\usepackage{hhline}
\usepackage[overload]{empheq}
\usepackage{cases}
\usepackage{nicefrac}
\usepackage{etoolbox}
\usepackage{multirow}
\usepackage{algorithm}
\usepackage{algorithmic}
\usepackage{subcaption}
\usepackage{cleveref}

\usetikzlibrary{plotmarks}
\usetikzlibrary{arrows.meta}
\usepackage{grffile}
\usepackage{amsmath}

\usepackage{dsfont}

\newcommand{\mybold}[1]{\bm{#1}}
\newcommand{\figref}[1]{Fig.~\ref{#1}}

\newcommand{\algref}[1]{Alg.~\ref{#1}}

\newcommand{\expv}{\mathbb{E}}

\DeclareMathOperator{\re}{Re}

\newcommand{\eye}[1]{\mybold{I}_{#1}}

\newtheoremstyle{remarkmod}
  {\topsep}   
  {\topsep}   
  {\normalfont}  
  {0pt}       
  {\itshape} 
  {.}         
  {5pt plus 1pt minus 1pt} 
  {}          
\theoremstyle{remarkmod}

\makeatletter
\newcommand*{\textoverline}[1]{$\overline{\hbox{#1}}\m@th$}
\makeatother

\tikzstyle{sum} = [draw, fill=blue!20, circle, node distance=1cm]
\tikzstyle{dot} = [draw, circle, minimum size=0.2pt,scale=0.3,fill=black,black]

\pgfdeclarelayer{background}
\pgfdeclarelayer{foreground}
\pgfsetlayers{background,main,foreground}

\newcounter{hints}
\renewcommand{\thehints}{\alph{hints}}
\newcommand{\hintedrel}[2][]{%
  \stepcounter{hints}%
  \if\relax\detokenize{#1}\relax\else\csxdef{hint@#1}{\thehints}\fi
  \mathrel{\overset{(\thehints)}{\vphantom{\le}{#2}}}%
}

\makeatletter
\newcommand*{\rom}[1]{\expandafter\@slowromancap\romannumeral #1@}
\makeatother

\makeatletter
\newcommand{\ALC@comblock}[1]{\ifthenelse{\equal{#1}{default}}%
{}{\textbf{#1}}}
\newenvironment{ALC@bl}{\begin{ALC@g}}{\end{ALC@g}}
\newcommand{\BLOCK}[2][default]{
	\ALC@it\ALC@comblock{#1}\ #2\begin{ALC@bl}
}
\newcommand{\ENDBLOCK}{
	\end{ALC@bl}
}
\newcommand{\PREDICT}{
	\BLOCK[Predict]
}
\newcommand{\ESTIMATE}{
	\BLOCK[Estimate]
}

\newcommand{\UPDATE}{
	\BLOCK[Update]
}

\newcommand{\GAIN}{
	\BLOCK[Kalman gain]
}

\makeatother

\newtoggle{showproof}
\toggletrue{showproof}
\begin{document}

	%
	%
	%

	%
	\maketitle
	\begin{abstract}
		Recently, channel state information (CSI) at the physical-layer has been utilized to detect spoofing attacks in wireless communications. However, due to hardware impairments and communication noise, the CSI cannot be estimated accurately, which significantly degrades the attack detection performance. Besides, the reliability of CSI based detection schemes is challenged by time-varying scenarios. To address these issues, we propose an adaptive Kalman based detection scheme. By utilizing the knowledge of the predicted channel we eliminate the channel estimation error, especially the random phase error which occurs due to the lack of synchronization between transmitter and receiver. Furthermore, we define a Kalman residual based test statistic for attack detection. Simulation results show that our proposed scheme makes the detection more robust at low signal-to-noise ratio (SNR) and in dynamic scenarios.
	\end{abstract}
	\begin{figure*}[b]
		\vspace{-0.3cm}
		\parbox{\hsize}{\em
			© 20XX IEEE. Personal use of this material is permitted. Permission from IEEE must be obtained for all other uses, in any current or future media, including reprinting/republishing this material for advertising or promotional purposes, creating new collective works, for resale or redistribution to servers or lists, or reuse of any copyrighted component of this work in other works.
	}\end{figure*}
	\section{Introduction}
	\label{sec:intro}
	
	The open nature of the radio propagation makes the wireless communication vulnerable to identity-based spoofing attacks, in which the attacker attempts to deliver a malicious message by pretending to be the legitimate user. In particular, by the communication over commodity WiFi networks, the attacker can simply use the command "ifconfig" to change its media access control (MAC) address and claim to be the authorized user \cite{pandey2012counter}. Therefore, the receiver must authenticate the message before proceeding with it. Traditional authentication mechanisms are based on encryption keys, which do not take into account the physical layer of the communication protocol. 
	In recent years,
	unique channel features extracted from the physical layer are exploited to enhance the authentication performance ~\cite{Authentication2015}. In practical communication protocols, such as the IEEE 802.11n~\cite{IEEEWiFi} standard, the orthogonal frequency-division multiplexing (OFDM) channel estimation mechanism is defined, with which the complex CSI can be obtained in discrete Fourier transform (DFT) domain. With the help of CSI the aforementioned spoofing attack can be effectively detected. 
	
	\noindent However, due to the lack of synchronization between transmitter and receiver, the CSI phase information is largely distorted. In more details, the time shift from the packet boundary detection results in packet detection delay (PDD) leading to random phase \emph{slope} error. Further, the carrier frequency offset (CFO) between the transmitter and receiver leads to random phase \emph{offset} error~\cite{ RCFOcompension,zhu2018splicer}. In many recent physical layer authentication studies, the problem is avoided by completely ignoring the observed CSI phase and focusing only on the received signal strength indicator (RSSI)~\cite{rumpel2016physical} or CSI magnitude~\cite{liu2018authenticating}. The conventional approach of CSI based detection schemes is by comparing the difference between the currently observed and the historical CSI~\cite{xiao2008using}. In recent years, machine learning (ML) based approaches have been developed in order to distinguish different transmitters, such as Gaussian mixture model (GMM)~\cite{Authentication2017,8468974} and support vector machine (SVM)~\cite{7924970,7037452}.    However, the wireless channel is time-varying. The stored historical CSI and the off-line learned channel features need to be updated over time, otherwise, it will cause great performance degradation.

	\noindent To solve these problems, in this paper we propose an adaptive Kalman filter based attack detection scheme that takes the predicted CSI into account. We formulate the attack detection process mathematically as a binary hypothesis testing problem. Unlike most state-of-art-studies that rely on historical CSI, we exploit the predicted CSI for attack detection. Furthermore, for attack detection we define a Kalman residual based test statistic, which follows a chi-squared distribution. The proposed scheme is evaluated by Monto Carlo simulations. Simulation results show that our proposed method outperforms most state-of-art attack detection schemes.

	\noindent The rest of the paper is organized as follows. The channel model is introduced in Sec.~\ref{sec:system}. We explain the proposed attack detection scheme in Sec.~\ref{sec:algorithm}.  Simulation results are given in Sec.~\ref{sec:results}.  The paper is concluded in Sec.~\ref{sec:conclusion}.

	\section{Channel model}
	\label{sec:system}
	
	Let $Q$ be the number of pilots used for channel estimation, $q_1,q_2,\ldots,q_Q\in\mathcal{Q}$ denote the pilot indices. Due to the communication noise and the synchronization problem the CSI can not be estimated accurately. We use a $Q \times 1$ vector ${\mybold{h}}_{\text{Obs},k}$ to denote the imperfect CSI estimation at time $k$, which can be expressed as 
	\begin{align}
	{\mybold{h}}_{\text{Obs},k} = \mybold{E}_{k}\mybold{C}{\mybold{h}}_{k} + \mybold{w}_{k},
	\label{eq:CSI}
	\end{align}
	where $\mybold{E}_{k}$ is a diagonal matrix that represents the phase error 
	\begin{align}
	\label{eq:diagonalPhase}
	\mybold{E}_{k} =  e^{j\Omega_{0,k}} 
	{\begin{bmatrix}
		e^{j\Omega_{d,k}q_1}, & & \\
		& \ddots & \\
		& &  e^{j\Omega_{d,k}q_Q} 
		\end{bmatrix}},
	\end{align}
	in which $\Omega_{0,k}$ and  $\Omega_{d,k}$ are the random phase distortion parameters caused by the CFO and PDD, respectively. Let $L$ be the channel length in time domain. The $Q \times L$ matrix $C$ is a partial DFT matrix with ${[C]}_{m,l}=e^{-j\frac{2\pi }{M}\cdot q_m\cdot l}$. $\mybold{h}_k$ is the channel in the time domain, which can be expressed as $
	\mybold{h}_k = {[h_{k}^{1},h_{k}^{2}, \dots,h_{k}^{L}]}^T
	$. We use $\mybold{h}_{\text{True},k}=\mybold{C} \mybold{h}_{k}$ to denote the "true" channel in DFT domain at time $k$. $\mybold{w}_{k}$ is the complex circularly-symmetric Gaussian noise with covariance $\sigma^{2}_{w}$.
	
	\noindent For the channel in time domain $\mybold{h}_k$, we assume a multi-path Rayleigh fading channel with Jakes doppler spectrum \cite{dent1993jakes} 
	\begin{align}
	{\Gamma}_{h^{(l)}}=\left\{\begin{matrix}
	\frac{{\sigma}_{h^{(l)}}^2}{\pi f_d\sqrt{1-{(\frac{f}{f_d})}^2}},& if \quad \left | f \right | < f_d\\ 0,&if\quad \left | f \right | \geq  f_d,	
	\end{matrix}\right.
	\end{align}
	where $f_d$ denotes the doppler frequency, ${\sigma}_{h^{(l)}}^2 =\expv[h_{k}^l{h_{k}^{lH}}] $ is the variance of the $l$-th channel path. In order to approximate the channel variations we apply here the first-order auto-regressive (AR1) 
	\begin{align}
	\mybold{h}_k = \alpha \mybold{h}_{k-1}+\mybold{v}_{k},
	\label{eq:cir}
	\end{align}
	where $\alpha$ denotes the channel correlation between previous time $k-1$ and $k$, $\mybold{v_k}$ is the circular-symmetric complex Gaussian process noise. According to Jakes spectrum, the transition parameter and the covariance matrix of $\mybold{v}_{k}$ can be obtained by the Yule-Walker equations \cite{baddour2005autoregressive}, which are given by
	\begin{align}
	\alpha = & J_0(2 \pi f_d T_s),\\
	\mybold{R}_k = & \expv[\mybold{v}_{k} \mybold{v}_{k}^{H} ] \\\nonumber = &(1-{\alpha}^2)diag([
	{\sigma}_{h^{(1)}}^2,{\sigma}_{h^{(2)}}^2,\cdots,{\sigma}_{h^{(L)}}^2]),
	\label{eq:sigma}
	\end{align} 
	where $f_d T_s$ is the normalized Doppler frequency, $J_0(\cdot)$ is the zero-order Bessel function. Here, $diag(\mybold{x})$ denotes creating a diagonal matrix whose main diagonal are elements of $\mybold{x}$.
	
	
	\section{Attack detection scheme}
	\label{sec:algorithm}
\begin{figure}
		\begin{tikzpicture}
		\node (ap) at (0,0) [inner sep=0em, label={[label distance=0em, text width=3.5cm, align=center]270:Bob}] {\includegraphics[width=1cm,keepaspectratio]{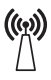}};
		
		\node (client1) [inner sep=0, label={[label distance=0em,text width=3.5cm, align=center]270:Alice}] at ($(ap)+(5,0)$) {\includegraphics[width=1cm,keepaspectratio]{Tx.png}};
		\node (jammer) [inner sep=0, label={[label distance=0em]270:Eve}] at ($(ap)+(3,-2)$) {\includegraphics[width=1cm,keepaspectratio]{Tx.png}};
		

		\draw[->,line width=1.5pt,blue] ($(client1)+(-0.5,0)$) to node[below, text width=2cm, align=center, yshift=0cm, pos=0.4] {${\mybold{h}}_{{\text{Obs},k}}^A$} ($(ap)+(1.5,0)$);
		\draw[->,line width=1.5pt,red,dotted] ($(jammer)+(-0.5,0.5)$) to node[below, text width=2cm, align=center, yshift=-0.35cm] {${\mybold{h}}_{{\text{Obs},k}}^E$} ($(ap)+(1,-0.25)$);
		\end{tikzpicture}
		\caption{Overview of the spoofing attack model}
		\label{fig:overview}
	\end{figure}

	We consider the spoofing attack model in~\figref{fig:overview}. A legitimate user (Alice) intends to communicate with Bob over the Alice-to-Bob channel. We use $\mybold{h}_{{\text{True},k}}^A$ and ${\mybold{h}}_{{\text{Obs},k}}^A$ to denote the true and the imperfect CSI of Alice-to-Bob channel, respectively. The attacker (Eve) tries to deliver a malicious message to Bob by pretending to be Alice. We use $\mybold{h}_{{\text{True},k}}^E$ and ${\mybold{h}}_{{\text{Obs},k}}^E$ to denote the true and the imperfect CSI of Eve-to-Bob channel, respectively. Bob has to decide whether the received message is from the legitimate user Alice or the attacker Eve. All users are assumed to be located in different positions with the location distance $d> \lambda $, where $\lambda$ is the radio frequency (RF) wavelength. Due to the \emph{location-specific} property of the wireless channel (${\mybold{h}}_{\text{True},k}^E \neq {\mybold{h}}_{\text{True},k}^A$), the CSI can be used to distinguish different transmitters.  
	However, due to lack of synchronization, the phase of the estimated CSI is largely distorted. Many state-of-art studies only focus on the magnitude of the CSI. In order to estimate the random phase error we have proposed an adaptive Kalman filter based algorithm in our previous work~\cite{chu2019phase}. In this work we apply the phase recovery approach for attack detection.

	\noindent We use $\hat{\mybold{h}}_{k|k-1}$ and $\hat{\mybold{h}}_{k|k}$ to denote the predicted and updated channel in the time domain, respectively. Furthermore, we use the $L\times L$ diagonal matrices $\mybold{P}_{k|k-1} $ and $\mybold{P}_{k|k}$ to denote predicted and the updated estimation covariance. Based on the state-space model defined in ~\eqref{eq:CSI} and~\eqref{eq:cir} we derive the adaptive Kalman process to jointly estimate the phase distortion parameters and the true channel in~\algref{alg:kalman}. Details can be found in~\cite{chu2019phase}. \begin{algorithm}
		\caption{Kalman filter based channel estimation}
		\label{alg:kalman}
		\begin{algorithmic}
			\REQUIRE Initialization of $\hat{\mybold{h}}_{0|0}$ and $\mybold{P}_{0|0}$
			\REPEAT
			\STATE $k \leftarrow k + 1$
			\PREDICT{channel estimate}
			\STATE $\hat{\mybold{h}}_{k|k-1}=\alpha \hat{\mybold{h}}_{k-1|k-1}$
			\ENDBLOCK
			\PREDICT{channel estimation error covariance}
			\STATE
			$\mybold{P}_{k|k-1} = {\alpha}^2 \mybold{P}_{k-1|k-1}+\mybold{R}_{k}$
			\ENDBLOCK
			\ESTIMATE{phase slope $\hat{\Omega}_{d,k}$ and phase offset $\hat{\Omega}_{0,k}$ }
			\STATE
			$\mybold{B}_k=  \mybold{E}({\Omega}_{0,k},{\Omega}_{d,k}) \mybold{C}$\\
			$\mybold{\Sigma}^{-1}=\left(
			\mybold{B}_{k}
			\mybold{P}_{k|k-1}
			\mybold{B}^{H}_{k}+
			\sigma^2_{w}\eye{Q}
			\right)^{-1}$\\
			$g=\mybold{h}^H_{\text{Obs},k}\mybold{\Sigma}^{-1}\mybold{h}_{\text{Obs},k} - 2\re\left[\mybold{h}^H_{\text{Obs},k}\mybold{B_k}\hat{\mybold{h}}_{k|k-1} \right]  $\\
			$
			\label{eq:loglikelihood2}
			\left \langle   \hat{\Omega}_{d,k},\hat{\Omega}_{0,k}\right \rangle =\underset{{\Omega}_{d,k},{\Omega}_{0,k}}{\arg\min} \, g\left( \Omega_{d,k},{\Omega}_{0,k}\right)$ \\
			
			\ENDBLOCK
			\GAIN{}
			\STATE
			$\mybold{K}_{k} =
			\mybold{P}_{k|k-1}
			\mybold{B}^{H}_{k}
			\left(
			\mybold{B}_{k}
			\mybold{P}_{k|k-1}
			\mybold{B}^{H}_{k}+
			\sigma^2_{w}\eye{Q}
			\right)^{-1} 
			$
			\ENDBLOCK
			\UPDATE{channel estimate}
			\STATE $\hat{\mybold{h}}_{k|k}=
			\hat{\mybold{h}}_{k|k-1}+
			\mybold{K}_{k}
			\left(
			\mybold{h}_{\text{Obs},k} - \mybold{B}_{k}\hat{\mybold{h}}_{k|k-1}
			\right)$
			\ENDBLOCK
			\UPDATE{channel estimation error covariance}
			\STATE
			$\mybold{P}_{k|k} = 
			\left(
			\eye{Q}-\mybold{K}_{k}\mybold{B}_{k}
			\right)
			\mybold{P}_{k|k-1}$
			\ENDBLOCK
			\UNTIL{forever}
		\end{algorithmic}
	\end{algorithm}   According to~\algref{alg:kalman}, we define the Kalman residual as
	\begin{align}
	\mybold{\epsilon}_k = \mybold{h}_{\text{Obs},k} - \mybold{B}_{k}\hat{\mybold{h}}_{k|k-1},
	\end{align}
	where $\mybold{B}_k=  \mybold{E}(\hat{\Omega}_{0,k},\hat{\Omega}_{d,k}) \mybold{C}$ is introduced for convenience, in which $(\hat{\Omega}_{0,k}$, $\hat{\Omega}_{d,k})$ denote the estimated phase distortion parameters. In the absence of attacks, the Kalman residual $\mybold{\epsilon}_k$ follows a complex Gaussian distribution with zero-mean and covariance matrix
	\begin{align}
	{\mybold{\Sigma}}_k = \underbrace{\mybold{B}_{k}
		\mybold{P}_{k|k-1}
		\mybold{B}^{H}_{k}}_{a}+
	\underbrace{\sigma^2_w\eye{Q}}_{b},
	\end{align}
	in which the terms $a$ and $b$ represent the channel estimation covariance in DFT domain and the covariance of the Gaussian noise $\mybold{\omega}_k$ according to the model defined in~\eqref{eq:CSI}, respectively. For the simplicity of the analysis, we assume here that the phase distortion parameters are estimated accurately. 
	Furthermore, we define the test statistic as
	\begin{align}
	\label{eq:teststatistic}
	{\lambda}_k =2{\mybold{\epsilon}_k}^H {{\mybold{\Sigma}}_k}^{-1}{\mybold{\epsilon}_k},
	\end{align}   
	which follows a chi-squared distribution with $2Q$ degree of freedom (DoF) in the absence of attacks. The chi-squared distribution can be expressed as
	\begin{align}
	{\lambda}_k \sim {{\chi}^{2}_{2Q}}.
	\end{align}    
	Thus, the threshold $d$ for attack detection can be evaluated for a given false alarm rate $P_{FA}$, which is given by 
	\begin{align}
	\label{eq:cdf}
	d(P_{FA}) = F^{-1}[1-P_{FA}|2Q]=\left\lbrace x:F(x|2Q)=1-P_{FA}\right\rbrace ,
	\end{align}
	where $F^{-1}$ denotes the inverse of the cumulative distribution function (cdf) of ${{\chi}^{2}_{2Q}}$. The cdf $F$ can be expressed as
	\begin{align}
	F(x|2Q) = \int_{0}^{x} \frac{t^{(2Q-2)/2}e^{-t/2}} {2^{Q} \Gamma (Q)}dt,
	\end{align}
	in which $\Gamma (\cdot)$ is the Gamma function. Thus, we formulate here the spoofing attack detection procedure as a binary hypothesis testing, which is given by 
	\begin{align}
	\mathit{H}_{0}:{\lambda}_k   \leqslant d(P_{FA}),  \\
	\mathit{H}_{1}:{\lambda}_k  > d(P_{FA}),
	\end{align} 
	where the null hypothesis $\mathit{H}_{0}$ denotes that the proposed test statistic is equal to or smaller than the threshold calculated at a given $P_{FA}$. This means that the received message at time $k$ is considered to be from the legitimate user Alice, while the alternative hypothesis $\mathit{H}_{1}$ denotes the received message is considered to be from the attacker Eve.
	Our proposed scheme is summarized as follows. At time $k$ the receiver Bob observes an imperfect channel estimate $\mybold{h}_{\text{obs},k}$ from an unknown transmitter. In order to eliminate the random phase errors, the Kalman prediction and phase estimation will be performed first. After that the receiver Bob calculates the test statistic ${\lambda}_k$ according to ~\eqref{eq:teststatistic}. If the test statistic is equal to or smaller than the threshold, Bob will accept the message and perform a Kalman update. Otherwise an alarm will generated and the current received message will be rejected.

	\section{Simulation results}
	\label{sec:results}
	\begin{figure}
		\centering
		\includegraphics[width=0.8\linewidth]{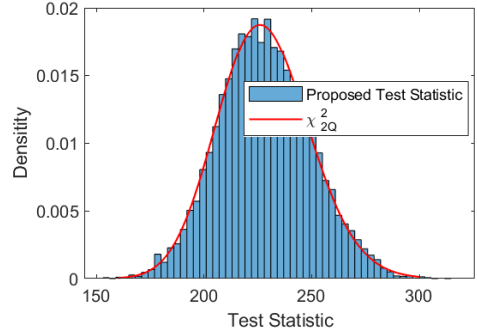}
		\caption{Test statistic distribution of ${\lambda}_{k}^{A}$, SNR = 10 dB, $f_d T_s = 10^{-4}$}
		\label{fig:teststatistic}
	\end{figure}
	\begin{figure}
		\centering
		\includegraphics[width=0.85\linewidth]{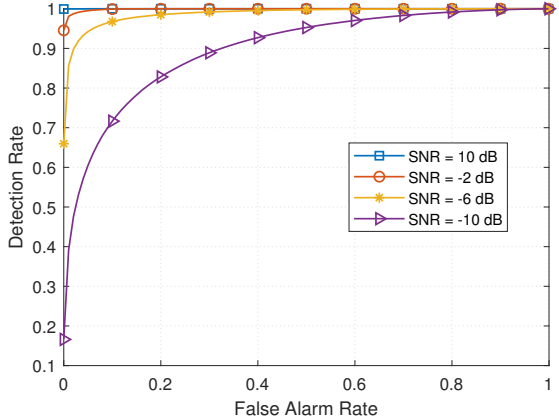}
		\caption{ROC of the proposed scheme, $f_d T_s=10^{-4}$}
		\label{fig:resultsproposed}
		
	\end{figure}
	
	
	\begin{figure*}
		\centering
		\includegraphics[width=0.9\linewidth]{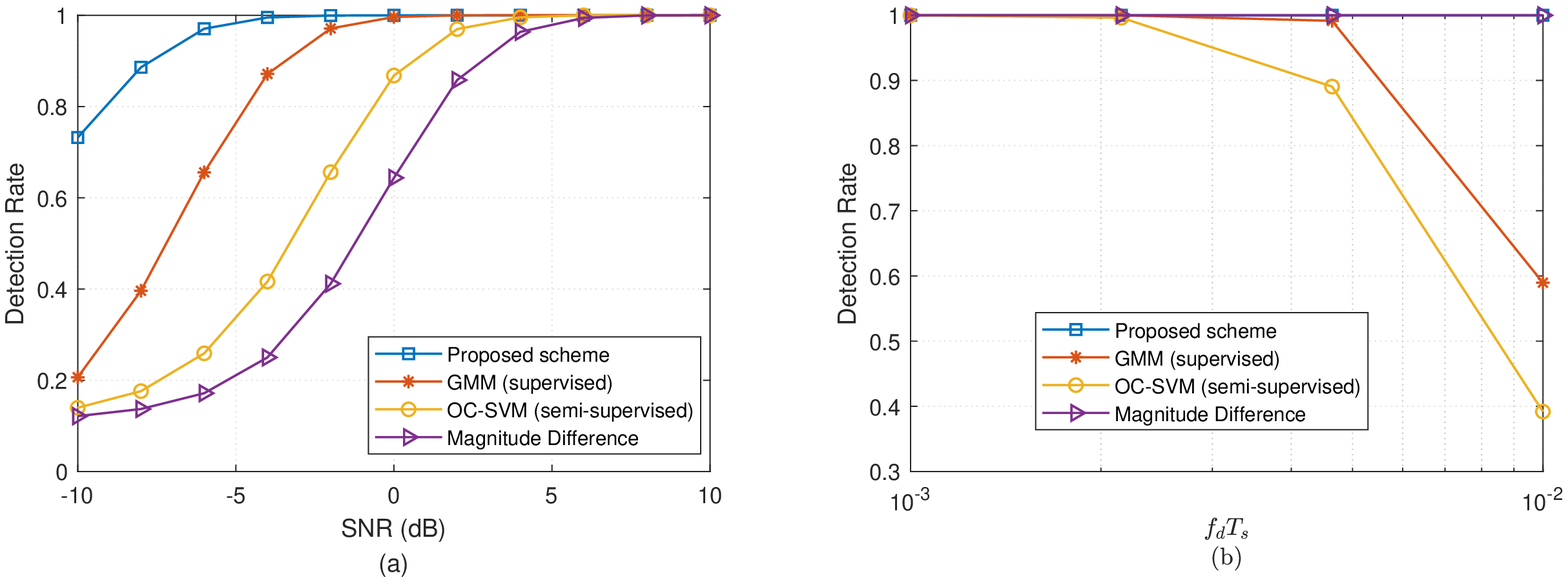}
		\caption{Comparison of different detection approaches: (a) Detection rate versus SNR, false alarm rate = 0.1, $f_d T_s = 10^{-4}$; (b) Detection rate versus $f_d T_s$, false alarm rate = 0.1, SNR = 10 dB  }
		\label{fig:results3}
		
	\end{figure*}
	%
	
	In this section we present the numerical results. A Monte Carlo simulation is performed to verify the
	proposed Kalman residual based attack detection. The results are averaged over $10^{4}$ simulations. According to the IEEE 802.11n standard\cite{IEEEWiFi}, we consider a OFDM system with 114 pilots. For all instances of the simulations, the channel in time domain is modelled as the multi-path Rayleigh channel with Jakes doppler spectrum. Meanwhile, the imperfect CSI is generated with complex Gaussian noise and random phase errors according to ~\eqref{eq:CSI}.  In order to perform the hypothesis testing, we generate 2000 imperfect CSI realizations of Alice-Bob channel (${\mybold{h}}_{{\text{Obs},k}}^A$) and Eve-Bob channel  (${\mybold{h}}_{{\text{Obs},k}}^E$) for each simulation. The entire Kalman filter based channel state recovery in~\algref{alg:kalman} is performed only for  ${\mybold{h}}_{{\text{Obs},k}}^A$ ($k=1,...,2000$). To evaluate the detection performance, after the prediction step we separately obtain the phase distortion terms given ${\mybold{h}}_{{\text{Obs},k}}^A$ and   ${\mybold{h}}_{{\text{Obs},k}}^E$. Then, according ~\eqref{eq:teststatistic} we calculate the proposed test statistics ${\lambda}_{k}^{A}$ and ${\lambda}_{k}^{E}$  using ${\mybold{h}}_{{\text{Obs},k}}^A$ and ${\mybold{h}}_{{\text{Obs},k}}^E$, respectively. 
	
	\noindent Theoretically, the proposed test statistics of Alice-Bob channel ${\lambda}_{k}^{A}$ should follow the chi-squared distribution. This is
	verified in~\figref{fig:teststatistic}. It can be clearly seen that the distribution of the test statistic obeys the chi-squared distribution well. The receiver operating characteristics (ROC) curves of the proposed scheme with different SNR are illustrated in~\figref{fig:resultsproposed}, in which each data point is a pair of detection rate and false alarm rate at a deterministic threshold. The threshold is calculated with a known false alarm rate according to~\eqref{eq:cdf}. The ROC curve represents the trade-off between the false alarm rate and the detection rate. From~\figref{fig:resultsproposed} we observe that as SNR increases, better detection performance can be achieved.
	
	\noindent We compare the proposed scheme to the approaches using GMM in~\cite{Authentication2017}, one class SVM (OC-SVM) in~\cite{7037452} and the magnitude difference between consecutive CSI in~\cite{7733691}. Note that, except of our proposed scheme the remaining approaches here only utilize the magnitude of the CSI, because the CSI phase is distorted severely due to the random errors. In addition, since GMM is a supervised ML based algorithm, we use the magnitude of ${\mybold{h}}_{{\text{Obs},k}}^A$ ($k=1,..,1000$) and   ${\mybold{h}}_{{\text{Obs},k}}^E$ ($k=1,..,1000$) to train the Gaussian mixture components, while the magnitude of ${\mybold{h}}_{{\text{Obs},k}}^A$ ($k=1001,..,2000$) and ${\mybold{h}}_{{\text{Obs},k}}^E$ ($k=1001,..,2000$) are used for testing. For the semi-supervised ML based OC-SVM approach, the magnitude of ${\mybold{h}}_{{\text{Obs},k}}^A$ ($k=1,..,1000$) are used for training, while the magnitude of ${\mybold{h}}_{{\text{Obs},k}}^A$ ($k=1001,..,2000$) and  ${\mybold{h}}_{{\text{Obs},k}}^E$ ($k=1001,..,2000$) are used for testing. Meanwhile we illustrate here the detection rate of the proposed scheme with $k=1001,..,2000$ for a fair comparison. In~\figref{fig:results3} (a), the detection rate is presented as a function of the SNR. It can be seen that our proposed scheme is superior to other methods, especially in the case of low SNR. The reason is that, through the Kalman filter based channel estimation in~\algref{alg:kalman}, we recover the CSI by the low-dimensional channel impulse response $\hat{\mybold{h}}_{k|k}$, thereby reducing the noise corruption. Additionally, we utilize the complex valued CSI, while the other approaches only using the magnitude. When we study the performance of the approaches with different doppler frequency as shown in~\figref{fig:results3} (b), we can see that the magnitude difference based approach performs similar to the proposed scheme.  The detection performance of ML-based algorithms decreases with higher Doppler frequencies, because the trained model for attack detection becomes obsolete due to channel variation.   
	\section{Conclusion}
	\label{sec:conclusion}
	In this paper, we have proposed a Kalman filter based spoofing attack detection scheme
	for dynamic channels. The detection problem is formulated as a binary hypothesis testing process with the defined test statistic, in which the predicted channel and the estimated phase errors are utilized. Simulation results have demonstrated that our proposed scheme outperforms most state-of-art approaches, especially in the case of low SNR and in dynamic scenarios. 
	\bibliographystyle{IEEEbib}
	\bibliography{refs}
\end{document}